\documentclass{moriond}

\usepackage[numbers]{natbib}
\usepackage{rotating}
\usepackage{graphicx}
\usepackage[utf8]{inputenc}
\usepackage{siunitx} 
\usepackage{gensymb}
\usepackage{amssymb}
\usepackage{amsmath}
\usepackage{color}

\newcommand{\fig}[1] {Figure~\ref{fig:#1}}

\newcommand{\aegis}{AE$\bar{\hbox{g}}$IS}
\newcommand{\Hbar}{$\bar{\hbox{H}}$}
\newcommand{\pbar}{$\bar{\hbox{p}}$}
\newcommand{\pos}{$\mathrm{e^+}$}

\newcommand{\Photo}{}

\DeclareSIUnit\inch{in.}
\DeclareSIUnit\division{div}

\begin{document}
\vspace*{4cm}

\title{\aegis{}: Status and Prospects}
\author{R. Caravita on behalf of the AEgIS collaboration}
\address{Trento Institute for Fundamental Physics and Applications, \\ 
via Sommarive 14, 38123 Povo (Tn), Italy}
\maketitle

\abstracts{
The progresses of the \aegis{} collaboration on its way towards directly measuring the gravitational free-fall of neutral antimatter atoms are reviewed. The experiment recently developed the first pulsed cold antihydrogen source and entered in its second phase, aiming at the first proof-of-concept gravitational measurement. Several major upgrades were deployed, including an upgraded antihydrogen production scheme and a fully-redesigned antiproton trap. \aegis{} re-started its operation on the new CERN ELENA decelerator in late 2021, capturing its first antiprotons and commissioning its new antiproton energy degrading system and hardware/software control systems.  
}

\section{Introduction}

The main goal of the \aegis{} experiment at CERN's Antimatter Factory is to measure the gravitational acceleration of antimatter atoms with a pulsed beam of cold antihydrogen (\Hbar{}) \cite{aegis_exp:15}. The method chosen by the collaboration is that of directly sampling the trajectory of free-falling atoms both in space and time, so to invert their (newtonian) equation of motion $ \Delta y \propto \bar{g} \, \Delta t^2 $ to work out $ \bar{g} $, Earth's gravitational acceleration experienced by antihydrogen atoms. Time-controlled antihydrogen production is realized by a charge-exchange process between trapped antiprotons (\pbar{}) and laser-excited Rydberg Positronium (Ps) atoms, whose cross-section is $ \sigma_{CE} \approx 10^{-15} \, {n_{Ps}}^4 \,\, \si{\centi\meter\squared}$ and $n_{Ps}$ is the principal quantum number of Rydberg Ps (Ps*, from now on)  \cite{krasnicky_pra:16}. Ps atoms are obtained by implanting bursts of \si{\kilo\electronvolt} positrons (\pos{}) in a nanochanneled mesoporous \pos{}-Ps silica converter \cite{aegis_targets:21}. 

\section{Review of \aegis{} phase 1 results}

\begin{figure}	[htp]
	\centering
	\includegraphics[width=0.48 \linewidth]{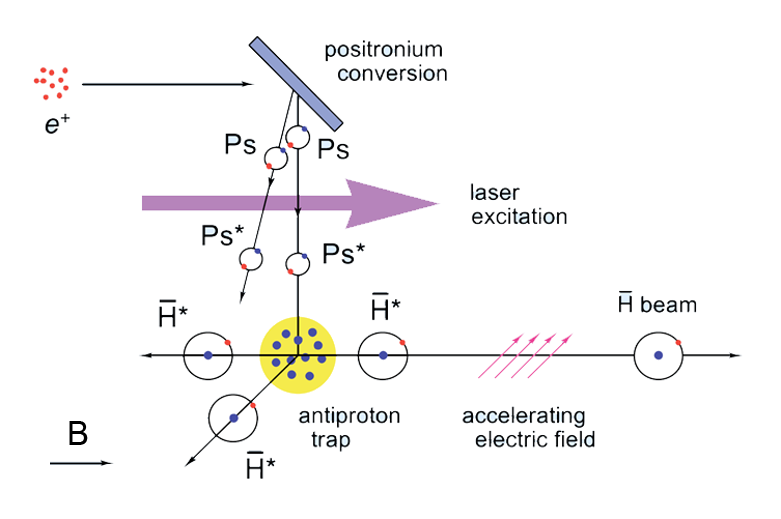}
	\includegraphics[width=0.48 \linewidth]{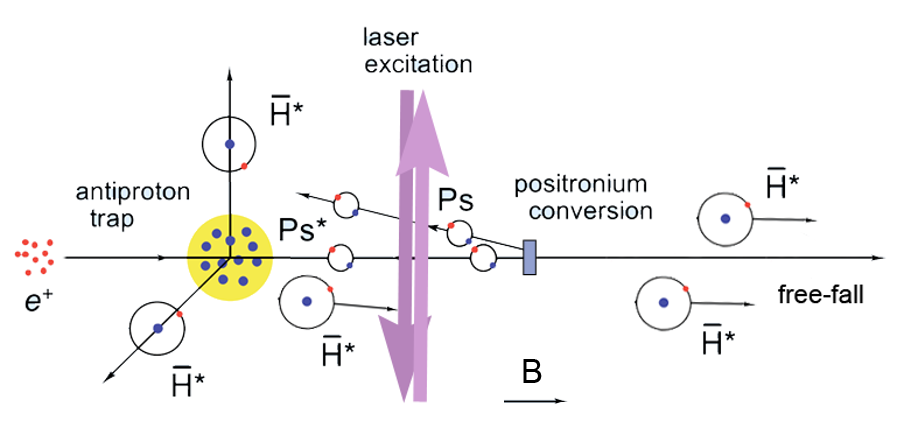}
	\caption{Left: orthogonal \Hbar{} production scheme of \aegis{} Phase 1: here Rydberg Ps atoms (Ps*) propagate on average orthogonal to the B field orientation, incurring in the maximal motional Stark effect (see text). Right: collinear \Hbar{} production scheme of \aegis{} Phase 2: here the Ps* atoms propagate on average parallel to the B field, cancelling the motional Stark effect.}
	\label{fig:aegis_scheme}
\end{figure}

During the first phase of the project (lasting approximately until the end of 2018, see \fig{physics_outlook}) \aegis{} aimed at three main objectives: assess experimentally the effectiveness of its selected gravitational sensor, establish Ps* production in its \SI{1}{\tesla} magnetic field environment, and use it to first produce \Hbar{} in a pulsed fashion following its original charge-exchange scheme (see \cite{aegis_exp:15}, also shown in \fig{aegis_scheme}, left).  
\\
First, the possibility to adopt a moiré deflectometer (the classical counterpart of a Mach-Zehnder interferometer) was investigated. \si{\kilo\electronvolt} \pbar{} were let go through a set of two \SI{40}{\micro\meter}-period gratings and detected by a nuclear emulsion detector \cite{aegis_emulsions2:13}. Comparing the so-obtained \pbar{} fringe pattern with a light reference, it was established that the device preserved its alignment in the cryogenic vacuum of the experiment, and that forces down to the \si{\atto\newton} sensitivity could be measured on a baseline of roughly \SI{20}{\centi\meter} \cite{aegis_natc:14}. 
\\ 
Second, Rydberg Ps production via two-step laser excitation in the \SI{1}{\tesla} magnetic $\vec{B}$ field of the experiment was realized. The maximum $ n_{Ps} $ that could be reached was $ n_{Ps} = 18 $ as, for higher levels, significant self-ionization of the atoms was observed. This resulted from the effect of an effective electric field experienced by the Ps* atoms in their co-moving frame of reference at velocity $\vec{v}_{Ps}$: the so-called \textit{motional Stark effect}, or $ \vec{E}_{MS} = \vec{v}_{Ps} \times \vec{B} $. This electric field caused efficient field-ionization of Ps* higher than a certain threshold: the maximum excitable $ n_{Ps} $, corresponding to a field-ionization rate of $\approx $ \SI{1}{\giga\hertz}, was $ n_{Ps} \approx 600 \, (9 \, v_{Ps} \, B \sin\theta)^{-1/4} \approx 19 $, where $ \theta = \pi/2$ is the trajectory angle between the individual Ps* trajectories and the magnetic field. \cite{aegis_velocimetry:20}.
\\
Last, a first pulsed cold \Hbar{} source was realized by letting $ n_{Ps} = 17 $ Ps* atoms impinge on a cloud of $ \approx 10^6 $ \pbar{} at \SI{400}{\kelvin}, held in a Malmberg-Penning trap with an entrance grid in the upper part of the central electrode to let Ps* fly through. \Hbar{} formation was established by comparing \pbar{} annihilation rates in the few \si{\micro\second} after Ps* production with and without the presence of antiprotons, positrons and the excitation laser. A significant excess of annihilations when both antiprotons, positrons and laser are enabled was observed, compatible with a production rate of $ 0.05 $ \Hbar{} atoms per cycle of the experiment (\SI{110}{\second}) \cite{aegis_hbar:21}.

\section{\aegis{} Phase 2}

\begin{figure}[htp]
	\centering
	\includegraphics[width=0.8 \linewidth]{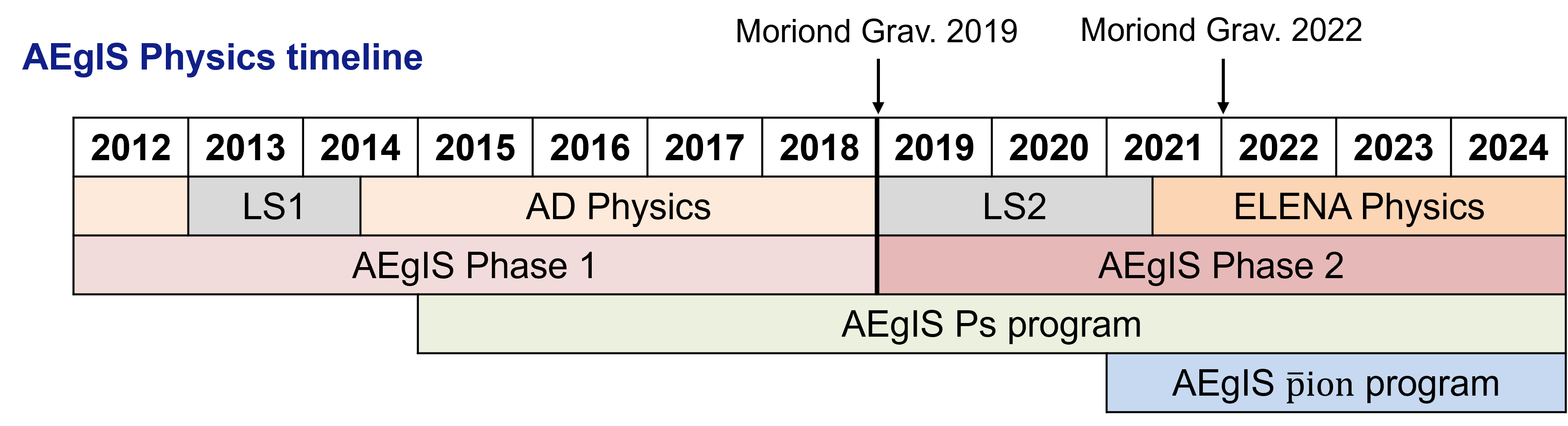}
	\caption{Outlook on the \aegis{} physics programs, underpinning the CERN decelerator schedules (first row), the leading antihydrogen physics program with its Phase 1 and Phase 2 (second row), and the seconday positronium and antiprotonic ions physics programs (third and fourth row).}
	\label{fig:physics_outlook}
\end{figure}

The success in realizing a first $ \approx $\SI{100}{\nano\second}-pulsed cold \Hbar{} source motivated the \aegis{} collaboration to continue its \Hbar{} physics program for a second major phase of the experiment (\fig{physics_outlook}), whose main goal is to achieve the first proof-of-concept inertial measurement with pulsed \Hbar{}. The main limitations of the \aegis{} phase 1 approach were identified to be the following: the \Hbar{} source intensity has to be increased by $2 \div 3$ orders of magnitude; the temperature of the \pbar{} plasma/produced \Hbar{} atoms has to be reduced by 1 order of magnitude; the experiment should be designed for the free-fall to take place in the most homogeneous magnetic field region of the experiment to limit systematic effects due to the magnet fringe-field. These sub-goals, together with the experience gained in producing and exciting Ps* in high magnetic fields, led to a major revision of the experimental scheme used to form \Hbar{}.  

A collinear production scheme (shown in \fig{aegis_scheme}, right) is adopted for \aegis{} phase 2, where the \pos{}-Ps silica converter target is installed on the same axis as that of the trap. This scheme offers several advantages, at the modest price of a limited opening loss in transmission. First and foremost, the motional Stark effect is cancelled to first order in the trajectory angle $ \theta $, as on average the Ps* atoms' distribution has $ \langle \theta \rangle \approx 0 $. This allows increasing $ n_{Ps} $ and excite Ps to Rydberg levels as high as $ n_{Ps} = 32 $, being limited only by the second-order angular spread of the distribution to cover efficiently the \pbar{} plasma in the trap: a cross-section gain of a factor of 16 is thus realized. Secondly, an up-to 50-fold increase in \pbar{} numbers is expected from the reduced \pbar{} energy of the ELENA accelerator with respect to the former Antiproton Decelerator, leading to a much higher trapping efficiency \cite{aegis_elena:18}. Combining these two enhancements, the \Hbar{} production rate will be brought to $1 \div 10 $ atoms for every experiment cycle. Lastly, the final trap of the experiment was fully redesigned: to reduce the number of electrodes and liberate $ \approx \SI{20}{\centi\meter} $ of flight space for the deflectometer; to minimize the RF noise footprint on the electrodes; to enforce a fully cylindrical symmetry and enhance the plasma stability; to embed an active alignment system at cryogenic temperatures to accurately align its axis to that of the magnetic field. These upgrades are meant to reach the state-of-the-art plasma temperatures in traps of $ \approx $ tens of K. 

\section{First antiprotons from ELENA}

\begin{figure}	[htp]
	\centering
	\includegraphics[width=\linewidth]{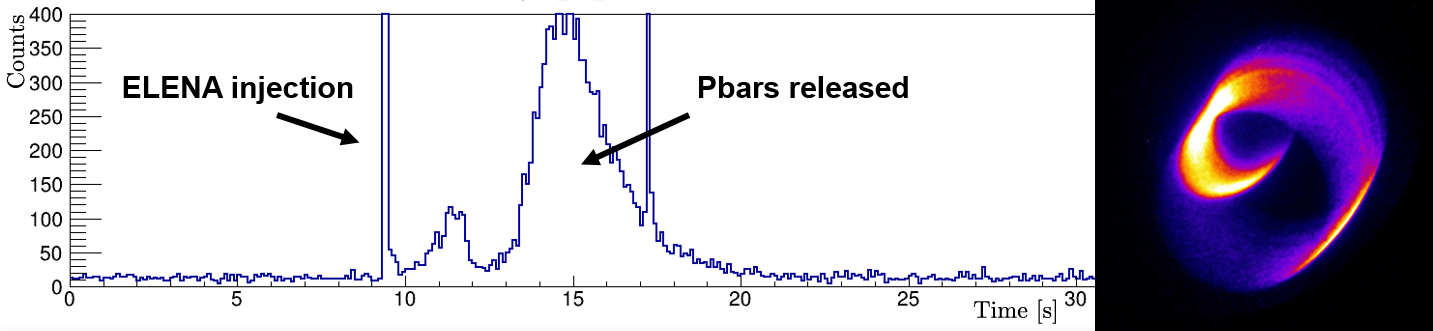}
	\caption{Left: first antiprotons caught by \aegis{} from ELENA during the 2022 run. Antiprotons were held in the trap for \SI{5}{\second} and released by turning off one of the two high-voltage electrodes used for capture, whose potential decreases in about \SI{5}{\second}. Right: a micro-channel-plate (MCP) image of the primary ELENA beam, chosen to emphasize the imaging capabilities of this detector to \SI{100}{\kilo\electronvolt} antiproton annihilations.}
	\label{fig:pbar_2022}
\end{figure}

\aegis{} restarted operation on October 20th, 2021, date on which the experiment took its first \pbar{} beam from the ELENA decelerator, equipped with a fully upgraded trap electronics and control system. This was a major upgrade step performed in 2020 (despite the difficulties imposed by the ongoing COVID-19 pandemic), consisting in the transition from a fully custom solution to a semi-standard LabVIEW/ARTIQ/Sinara software/hardware ecosystem, featuring the high flexibility of the Python-like programming interface of ARTIQ, combined with the \si{\nano\second}-scale hardware synchronization capability of the Sinara hardware and with the solidity of a distributed LabVIEW control system based on the Actor Model of concurrency. 
\\
The 2021 run allowed \aegis{} to test its upgraded control system, upgraded energy degraders and detection system. Antiprotons from ELENA were successfully imaged and steered on a micro-channel plate detector (MCP) (\fig{pbar_2022}, right), which allowed careful determination of the optimal beam steering for energy degrading. The energy degrading effect from its overall \SI{1600}{\nano\meter} thickness of Parylene N foils was observed, with $ \approx 15 \% $ of the energy distribution of \pbar{} transmitted by the degrader lower than \SI{10}{\kilo\electronvolt}, the threshold for trapping with the current high-voltage electrodes \cite{aegis_spsc:22}. The first trapping of \si{\kilo\electronvolt} \pbar{} was achieved, as shown in \fig{pbar_2022}, left. Trapped \pbar{} were held in the catching trap for a few seconds, before being slowly released and imaged by a scintillation detector by ramping down the high-voltage trapping electrodes in about \SI{5}{\second}.

\section{Outlook and future prospects}

\aegis{} has recently entered in its second phase (see \fig{physics_outlook}), after the successful demonstration of the first pulsed source of cold \Hbar{}, aiming at the first proof-of-concept gravitational measurement with antimatter atoms. In the near future, the collaboration aims at showing that an enhancement in the \Hbar{} atoms flux from its source by up to a factor of 100 is possible, as well as reducing the produced atoms' temperature. As the following step, \aegis{} aims at forward-boosting the produced \Hbar{} to $ \approx 10^3 \,\, \si{\meter\per\second} $, by collectively accelerating the \pbar{} plasma right before the charge-exchange occurs, to impress on the \pbar{} a common axial momentum that is then transferred to the \Hbar{} atoms. This will effectively realize a pulsed, forward-boosted beam of transversally-cold \Hbar{}, the necessary ingredient for attempting a first gravitational measurement, which is thus possible in the mid future. The parallel development of a suitable deflectometer/interferometer is already ongoing, to develop an instrument able to measure the tiny vertical deflection due to gravity: $ \delta y \approx 0.2 \, \si{\micro\meter} $ for $10^3 \,\, \si{\meter\per\second}$ \Hbar{} in the \SI{20}{\centi\meter} baseline available for the experiment. 

\section*{Acknowledgments}

This work was sponsored by the European’s Union Horizon 2020 research and innovation program under the Marie Sklodowska Curie grant agreement No. 754496, FELLINI, by the Wolfgang Gentner Programme of the German Federal Ministry of 
Education and Research (grant no. 05E18CHA  and by Warsaw University of Technology within the Excellence Initiative: Research University (IDUB) programme and the IDUB-POB-FWEiTE-1 project grant.

\nocite{*}
\bibstyle{unsrt}
\bibliography{the_biblio}

\end{document}